\documentclass[aps,jcp,preprint,floatfix,superscriptaddress]{revtex4-1}
\usepackage{graphicx}
\usepackage{dcolumn}
\usepackage{amsmath}
\usepackage{amsfonts}
\usepackage{color}

\begin{document}

\title{Equilibrium properties of charged microgels:\\ a Poisson-Boltzmann-Flory approach}

\author{Thiago Colla}
\email{thiago.colla@ufrgs.br}
\affiliation{Faculty of Physics, University of Vienna, Boltzmanngasse 5, A-1090 Vienna, Austria}

\author{Christos N. Likos}
\email{christos.likos@univie.ac.at}
\affiliation{Faculty of Physics, University of Vienna, Boltzmanngasse 5, A-1090 Vienna, Austria}

\author{Yan Levin}
\email{levin@if.ufrgs.br}
\affiliation{Instituto de F\'isica, Universidade Federal do Rio Grande do Sul, Caixa Postal 15051, CEP 91501-970, Porto Alegre, RS, Brazil}

\begin{abstract}

The equilibrium properties of ionic microgels are investigated using a combination of the Poisson-Boltzmann and Flory theories. Swelling behavior, density profiles, and effective charges are all calculated in a self-consistent way. Special attention is given to the effects of salinity on these quantities. In accordance with the traditional ideal Donnan equilibrium theory, it is found that the equilibrium microgel size is strongly influenced by the amount of added salt. Increasing the salt concentration leads to a considerable reduction of the microgel volume, which therefore releases its internal material -- solvent molecules and dissociated ions -- into the solution. Finally, the question of charge renormalization of ionic microgels in the context of the cell model is briefly addressed. 

\end{abstract}

\maketitle

\section{Introduction}

Cross-linked microgel particles are quite remarkable due to their large sensibility on the external conditions \cite{Flory,Christos,Likos2001}. The interactions among these particles are known to be strongly influenced by experimentally controlled quantities such as the temperature, the solvent quality, the particle concentration, the ionic strength or the degree of cross-linking, among many others  \cite{Likos2001,Escobedo1999,Riest2012}. Depending on the particular way these particles are synthesized, different effective interactions among them can be induced \cite{Antoni1990,Gottwald2004}. Another way to drive desirable effective interactions among these particles is by submitting them to controlled external fields \cite{Nojd2013}. The possibility of steering the dynamical and equilibrium properties of such particles through changes in the surrounding environment makes them promising in a number of chemical, biological as well as medical applications \cite{Bomberg2002,Eichenbaum1999,Antoni1990}. In contrast to most of the traditional hard colloidal systems, the soft nature of the short range interactions of microgels opens the possibility to generate systems with extremely high packing fractions \cite{Sessoms2009}. Another important characteristic that distinguishes microgels from hard colloidal particles is their permeability. Depending on the external conditions, solvent molecules can flow into or leave the microgels, resulting in a swelling (or de-swelling) of the cross-linked network. The fact that microgels can exchange particles with their environment makes them well suitable for their application in the design of drug-delivery mechanisms, where molecules can be encapsulated -- and further released -- in specific targets through this swelling process \cite{Bomberg2002,Antoni1990,Escobedo1999,Eichenbaum1999}. 

When in contact with an aqueous solvent, a fraction of monomers inside the microgels become dissociated, releasing their counterions into the bulk solution. The resulting system is then composed of microgels with their charged cross-linked polymer chains, solvent molecules and counterions, along with possible ions of dissociated salt \cite{Likos2001,Christos,Levin2002}. The presence of charged components strongly increases the system complexity. Apart from the long range nature of the Coulomb interactions, contributions from the charge balance due to the addition of salt -- the so-called Donnan equilibrium effects --  must be carefully considered \cite{Donnan1924,Tamashiro1998}. In addition, the presence of salt is known to have a non trivial influence in the underlying thermodynamics of charged systems \cite{Colla2012}. Due to the strong electrostatic interactions between counterions and the charged backbones, the majority of the former will remain trapped inside the microgels, while solvent molecules can flow freely through the microgel-solution interface. The equilibrium properties are then mostly dictated by the chemical equilibrium between these components across the interface \cite{Flory1943,Flory}, along with the elastic contributions from the cross-linked network.  

A theoretical description which takes into account the chemical and physical contributions in charged microgel systems in a detailed level is way too complex. In this context, simple approximate models which help to highlight the key physical mechanisms of the underlying phenomena prove to be extremely useful. When dealing with charged objects in the presence of monovalent ions in an aqueous environment, the mean field Poisson-Boltzmann (PB) theory provides a manageable description, yet with an excellent degree of accuracy \cite{Levin2002,Marcus1955,Deserno2001}. Contrary to most of the liquid-state integral equation theories, the PB equation allows for a transparent physical interpretation of complex phenomena involving charged components -- whenever the mean-field picture holds. It has been successfully applied to describe a variety of complex systems where macromolecules are surrounded by monovalent ions \cite{Levin2002,Deserno2001}. In the case of charged macroions which are permeable to the surrounded counterions, the PB formalism has been recently applied to study the ionic profiles, the effective charges as well as the charge renormalization of such systems \cite{Chepelianskii2009,Chepelianskii2011,Bauli2012}. When the penetrable particles are surrounded by divalent ions the mean field PB theory breaks down, and more sophisticated approaches have to be used to account for the strong electrostatic correlations \cite{Moncho-Jorda2014}.

In what concerns the swelling behavior of microgels, a number of experimental and theoretical works have been carried out over the years to elucidate the physical mechanisms behind this phenomenon, for both cases of charged \cite{Barrat1992,Levin_Diehl2002,Fernandez-Nieves2001,Fernandez-Nieves2003} and neutral \cite{Flory1943,Holmqvist2012,Zhi2010,Zhi2011,Huang2008,Routh2006,Eichenbaum1999,Sierra-Martin2012} microgels. For a recent review on this fascinating topic, we refer the reader to Ref.[\onlinecite{Quesada-Perez2011}]. In the case of ionic microgels, the effects of increasing the ionic strength and the polymer charge fraction over the swelling properties have been extensively investigated by means of computer simulations \cite{Mann2005,Yin2008,Yin2009,Quesada-Perez2012}, experiments \cite{English1996,Fan2010,Nerapursi2006,Nisato1998,Horkay2000,Xia2003,Dubrovskii1997,Lopez-Leon2006} and theory \cite{Pincus1991,Barrat1992,Katchalsky1955,Quesada-Perez2011,Rydzewski1990,Victorov2006,Yigit2012}. It is now well established that the increase in the salt concentration leads to the particle de-swelling, while the dissociation of polymer chains produces an increase in the microgel volume. These qualitative effects can be captured by the traditional Donnan theory for the ionic contributions to the microgel osmotic pressure \cite{Hooper1990,Horkay2000,Horkay2001,Fernandez-Nieves2001,Fernandez-Nieves2003,Barrat1992}. In this leading-order approximation, a chemical equilibrium between an electroneutral microgel and an infinite salt reservoir is assumed, and the osmotic pressure follows from the ideal gas ionic contributions only \cite{Barrat1992}. Together with the Flory elastic and mixing contributions, this approach provides a simple and transparent way to qualitatively account for the  swelling properties of ionic microgels. Several modifications have been proposed to investigate the swelling equilibrium beyond this simple approach \cite{Rydzewski1990,Katchalsky1955,Victorov2006,Capriles-Gonzalvez2008,Hoare2007,English1998,Polotsky2013}. An improvement over this classical ideal Donnan picture consists in introducing ionic correlations in a Debye-H{\"u}ckel (DH) level of approximation \cite{Rydzewski1990,Victorov2006}. As pointed out by English {\it et al} \cite{English1998}, even this linear DH approximation is not sufficient to correctly reproduce the experimentally observed ionic contributions to the particle swelling at high polymer charges and salt concentrations, in such a way that higher order terms have to be taken into account in the virial expansion.  Furthermore, it is expected that strong non-linear effects will take place close to the microgel surface, where the electrostatic potential undergoes an abrupt decay \cite{Barrat1992,Chepelianskii2009}. Obviously, these effects can not be captured by the linear DH theory. It is therefore not yet clear how these effects may influence the Donnan equilibrium across the interface. The aim of the present work is to provide a self-consistent theory that combines the aforementioned accuracy of the PB equation for strongly charged ionic microgels with the classical thermodynamic Flory theory for the microgel volume transitions. The main focus will be to determine how the ionic contributions influence the swelling behavior in the framework of the PB theory, as the ionic strength and the bare microgels charge are changed.

The remaining of the Paper is organized as follows. In section II, the system under consideration is described in some detail. The construction of the variational mean-field theory is made in section III, along with a description of its numerical implementation. The results for several equilibrium properties are presented in section IV, followed by discussion and conclusions in section V.

\section{The system} 

We consider a system of cross-linked microgel particles immersed in an aqueous environment at fixed room temperature. The microgels are made of $N$ flexible chains, each of which carries a number $m$ of spherical monomers of radius $r_{m}$. Due to the high solvent dielectric permittivity, a fraction $f$ of these monomers dissociates producing $Z=fNm$ anionic monomers and $Z=fNm$ cationic counterions. 
Besides microgels, strong 1:1 electrolyte (salt) at concentration $c_{s}$ is also present in the
solution.  Dissociation of salt leads to additional coions (anions) and counterions (cations) each at concentration $c_{s}$. For simplicity, we will assume that both ions and solvent molecules are spherical objects of radius $r_{i}$. 

\begin{figure}[h]
\centering
\includegraphics[scale=0.8]{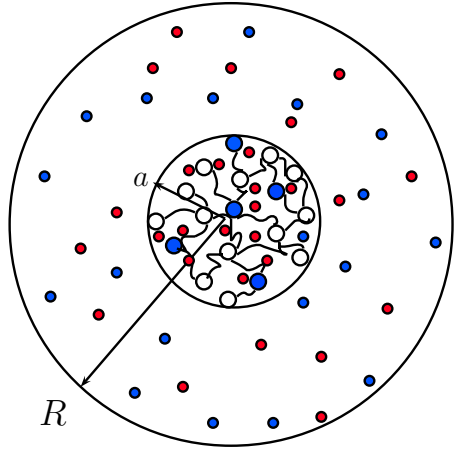}
\caption{Schematic representation of the system. A microgel of radius $a$ containing $N$ cross-linked chains, carrying $m$ monomers each (bigger spheres), is placed at the center of a spherical  WS cell. The radius $R$ of the cell is fixed by the overall microgel concentration $\rho$ inside the solution, $R=\left(\dfrac{3}{4\pi\rho}\right)^{1/3}$. A fraction $f$ of the monomers is dissociated (blue big spheres), releasing their counterions. Both counterions (small red spheres) and coions (small blue spheres) are free to move all over the WS cell volume. The solvent is represented by the background in which particles move.}
\label{fig0}
\end{figure}

Instead of considering the full microgel solution explicitly, we adopt a Wigner-Seitz (WS) cell model, in which a single microgel of radius $a$ is placed at the center of a spherical cell of radius $R$ (see Fig. \ref{fig0}). The cell is taken to be electrically neutral, and its radius $R$ is determined by the concentration of the microgels inside the solution, $4\pi R^3 /3=1/\rho$, where $\rho$ is the overall microgel concentration. Both microions and solvent molecules are free to move throughout the cell volume, while the fixed polymer backbones are confined in the interior of the microgel.

\section{The model}

We begin by constructing the total Helmholtz free energy inside the cell as a function of the microgel radius $a$, for a given salt concentration $c_{s}$ inside the WS cell, fraction of dissociation $f$ and density $\rho$. This Helmholtz free energy can be split into ionic, solvent and elastic contributions, $\beta\mathcal{F}=\beta\mathcal{F}_{ion}+\beta\mathcal{F}_{sol}+\beta\mathcal{F}_{el}$. We now turn to the calculation of each one of these terms separately.

\subsection{Ionic free energy}

The ionic free energy can be written as a functional of the ionic particle distribution inside the cell, 
$\rho_{\pm}(\mathbf{r})$. The inhomogeneity is provided by the interaction between the ions and the charged monomers lying inside the microgel. We adopt here a mean field description in which 
the electrostatic correlations and the exclusion volume effects are ignored. Furthermore, we will suppose that the polyelectrolyte network of the microgel provides a uniform charged background in which cations
and anions move.  The charge density  $\varrho_{m}(\mathbf{r})$ of the background is:
\begin{equation}
\varrho_{m}(\mathbf{r})=-\dfrac{3Zq}{4\pi a^3}\Theta(a-r),
\label{rhom}
\end{equation}
where $q$ is the charge of a proton and $\Theta(x)$ is the Heaviside step function. Within the mean-field  approximation, the ionic free energy can be written as
\begin{eqnarray}
\beta\mathcal{F}_{ion}[\rho_{\pm}(\mathbf{r})] =  \sum_{i=\pm}\int{\rho_{i}(\mathbf{r})\left[\ln\left(\lambda_{B}^3 \rho_{i}(\mathbf{r})\right)-1\right]d\mathbf{r}}+\dfrac{\beta }{2}\int{\left[\varrho_{+}(\mathbf{r})-\varrho_{-}(\mathbf{r})-\varrho_{m}(\mathbf{r})\right]\psi(\mathbf{r})d\mathbf{\mathbf{r}}} \nonumber  \\ -3\ln\left(\dfrac{\lambda}{\lambda_{B}}\right)\sum_{i=\pm}N_{i}, \hspace{5cm}  
\label{Fion1}
\end{eqnarray}
where $\varrho_{\pm}(\mathbf{r})\equiv q\rho_{\pm}(\mathbf{r})$ are the ionic charge distributions, $\lambda$ is the thermal de Broglie wavelength, $\lambda_{B}=\beta q^2/\epsilon$ is the Bjerrum length,  $\epsilon$ is the solvent dielectric constant and $\psi(\mathbf{r})$ is the mean electrostatic potential inside the WS cell. The first term on the right-hand side of this expression is the ideal-gas contribution of the mobile ions, while the second term represents the electrostatic energy and the third term is an irrelevant constant.  At this level of approximation, the ions are free to go into the microgel, resulting in a large counterion penetration. This equation has to be solved under the constraint of fixed number of coions and counterions $N_{\pm}$ inside the cell:
\begin{equation}
\int\rho_{\pm}(\mathbf{r})d\mathbf{r}=N_{\pm},
\label{RHOpm1}
\end{equation}
where $N_{-}=c_{s}V$ and $N_{+}=c_{s}V+Z$. In equilibrium, the density profiles should be the ones that minimize the functional $\mathcal{F}_{ion}[\rho_{\pm}(\mathbf{r})]$ subject to the conditions (\ref{RHOpm1}). Applying the minimization condition, one easily finds $\rho_{\pm}(\mathbf{r})=c_{\pm}e^{\mp\beta q\psi(\mathbf{r})}$, where $c_{\pm}\equiv \exp(\beta \mu_{\pm})/\lambda^3$, and $\mu_{\pm}$ are the Lagrange multiplayers necessary to satisfy Eq. (\ref{RHOpm1}). Together with the Poisson equation, this relation leads to the Poisson-Boltzmann equation for the mean electrostatic potential:
\begin{equation} 
\nabla^2 \phi(r)= -4\pi\lambda_{B}\left(c_{+} e^{-\phi(r)}-c_{-}e^{\phi(r)}-\dfrac{3Z}{4\pi a^3}\Theta(a-r)\right),
\label{PB1}
\end{equation}
where $\phi(r)\equiv\beta q \psi(r)$. This equation is numerically solved under the condition of charge neutrality inside the cell, $\phi'(R)=\phi'(r\rightarrow 0)=0$. Once the numerical solution is obtained, the corresponding ionic contribution to the free energy follows directly from the substitution of the ionic profiles $\rho_{\pm}(\mathbf{r})$ in Eq. (\ref{Fion1}). 

It is important to keep in mind that the solvent has been so far considered only implicitly, through its permeability $\epsilon$. Another relevant point to be stressed relies on the fact that the ionic distributions resulting from this variational procedure are \textit{not} the optimal density profiles for the {\it set of parameters considered}. This is because the ionic profiles obtained from Eq. (\ref{PB1}) have an implicit dependence on the microgel radius, $\rho_{\pm}(\mathbf{r})=\rho_{\pm}(\mathbf{r};a)$. Since the ionic contributions will influence the swelling process, the microgel radius $a$ is not known {\it a priori}, and has to be calculated in a self-consistent way. The {\it equilibrium} density distributions will be the ones for which the condition of minimization of the {\it total} free energy with respect to the particle size $a$ is satisfied, as described in the following sections. 

\subsection{Solvent free energy}

The solvent is modeled as a uniform background in which the ions move. Like the ionic species, the solvent molecules can go all the way to the interior of the microgel particle, resulting in its swelling. Neglecting solvent-ion and solvent-solvent interactions, the solvent contribution to the free energy can be written as a sum of entropic and solvent-polymer contributions, $\beta\mathcal{F}_{sol}=\beta\mathcal{F}_{id}+\beta\mathcal{F}_{sol-pol}$. The ideal gas contribution is
\begin{equation}
\beta\mathcal{F}_{id}=N_{s}^{in}(\ln(\phi_{s}^{in})-1)+N_{s}^{out}(\ln(\phi_{s}^{out})-1)+3N_{s}\ln\left[\dfrac{\lambda}{r_{i}}\right],
\label{Fid}
\end{equation}   
where $\phi_{s}^{in}$ and $\phi_{s}^{out}$ represent the solvent volume fraction inside and outside the microgel, respectively, $N_{s}^{in}$ and $N_{s}^{out}$ being the corresponding particle numbers. The last term of the right-hand size, involving the thermal de Broglie wavelength $\lambda$, represents here an irrelevant constant. For a {\it fixed} microgel radius $a$, these quantities can be easily expressed in terms of the number of ions condensed inside the microgel particle, $N_{cond}$. To this end, we assume that the ``empty" space -- which is neither occupied by ions nor by the polymer backbones -- is the total volume of the solvent background:
\begin{eqnarray}
\phi_{s}^{in}&=&\dfrac{a^3 -Nm r_{m}^3 -N_{cond} r_{i}^3}{a^3} \\  \\
\phi_{s}^{out}&=&\dfrac{R^3 - a^3 -(N_{+}+N_{-}-N_{cond})r_{i}^3}{R^3 - a^3}. \nonumber
\end{eqnarray}
Assuming the solvent molecules as spherical particles having the same size as the ionic species $r_{i}$, the number of such molecules inside and outside the microgels can be simply written as
\begin{eqnarray}
N_{s}^{in}&=&\left(\dfrac{a}{r_{i}}\right)^3 \phi_{s}^{in}, \\ \\
N_{s}^{out}&=&\left(\dfrac{R^3 - a^3}{r_{i}^3}\right) \phi_{s}^{out} \nonumber.
\end{eqnarray}
In order to construct the solvent entropic free energy, it only remains to calculate the number of ions which lie inside the microgel, $N_{cond}$. This quantity is self-consistently obtained once the density profiles $\rho_{\pm}(r)$ for the given microgel size $a$ have been calculated through the PB equation (\ref{PB1}):
\begin{equation}
N_{cond}=4\pi\int_{0}^{a}r^2 \left[\rho_{+}(r)+\rho_{-}(r)\right]dr.
\label{Ncond}
\end{equation}
Apart from the entropic contributions, the solvent free energy also contains the contribution from the interaction between the solvent and the hydrophobic polymer backbones inside the microgel. According to the mean-field Flory theory, this quantity is given by
\begin{equation}
\beta\mathcal{F}_{sol-poly}=Nm\chi\phi_{s}^{in},
\label{Fsol-poly}
\end{equation}
where $\chi$ is the Flory solvent-polymer parameter \cite{Flory}. Clearly, for a hydrophobic polymer backbone $\chi>0$ this contribution has the effect of repelling the solvent molecules from the interior of the microgel. 

Despite its implicit functional dependence on the ionic profiles via Eq. (\ref{Ncond}), it is important to note that the solvent free energy has been not considered in the functional minimization procedure that leads to Eq. (\ref{PB1}), where only the ionic contributions (coulombic interactions) were subject to minimization with respect to the density profiles $\rho_{\pm}(r)$. This is because size effects are here fully ignored at the \textit{functional} level of approximation. In the present description, there is no \textit{real} solvent-ion interactions, and the only ionic effect in the solvent contributions is through the ionic exclusion size, which effectively reduces the overall volume available for the solvent molecules to move in, leading to a  renormalization of its volume fractions. At the mean-field level of description that leads to the PB equation, ionic exclusion effects are completely neglected, and it is therefore fully consistent to also neglect ion-solvent size effects at the same (functional) level of approximation. This procedure is justified {\it a posteriori} by explicitly checking that in the limit of point-like ions ($r_{i}\rightarrow 0$) -- where the functional dependence in Eq. (\ref{Ncond}) vanishes and the present formalism becomes exact -- the results are qualitatively unchanged. On the other hand, a complete solvent description would require the construction (and mutual minimization) of a coupled solvent-ion density functional with inhomogeneous solvent distribution, as well as density-dependent solvent-polymer interactions \cite{Onuku2004,Okamoto2011,Bier2012}, which is beyond the scope of this work.  

\subsection{Elastic free energy}

Upon deformation, the microgels experience an elastic response as a result of the change in the conformation of their polymer chains. Assuming a microgel is isotropic, the elastic contribution to the free energy can be written as \cite{Levin2004,Flory}
\begin{equation}
\beta\mathcal{F}_{el}=\dfrac{3N}{2}(\alpha^2 -\ln\alpha-1),
\label{Fel}
\end{equation}
where $\alpha$ is the microgel expansion factor, which is proportional to the ratio between its actual volume $V$ and the volume in the unstressed state $V_{0}$:
\begin{equation}
\alpha=\left(\dfrac{V}{V_{0}}\right)^{1/3}=\dfrac{a}{(Nmr_{m}^3 + Z r_{i}^3)^{1/3}}.
\label{alpha}
\end{equation}
In the second equality of this expression, we have used the fact that the unstressed, equilibrium volume corresponds to the dry state where monomers and counterions are in their close-packed configuration.

\subsection{Equilibrium condition}

Once the equilibrium density profiles $\rho_{\pm}(\mathbf{r})$ are obtained through the solution of the PB equation, Eq. (\ref{PB1}), at \textit{fixed} microgel radius $a$, the total free energy inside the cell as a function of $a$ can be calculated by combining Eqs. (\ref{Fion1}), (\ref{Fid}), (\ref{Fsol-poly}) and (\ref{Fel}),
\begin{equation}
\beta\mathcal{F}(a)=\beta \mathcal{F}_{ion}+\beta\mathcal{F}_{id}+\beta\mathcal{F}_{sol-poly}+\beta\mathcal{F}_{el}.
\label{Ft}
\end{equation}

For a given number of chains $N$, number of monomers per chain $m$, salt concentration $c_{s}$ and a fraction of dissociation $f$, the equilibrium state will be determined by the minimization of the free energy with respect to the microgel radius, keeping constant all the remaining system parameters:
\begin{equation}
\dfrac{\partial \beta \mathcal{F}}{\partial a}\biggr\arrowvert_{N,m,c_s,f}=0.
\label{equ}
\end{equation}
This condition is equivalent to the mechanical requirement that the internal microgel pressure must be exactly balanced by the external one across the microgel-solution interface. 

The equilibrium microgel size as determined by Eq. (\ref{equ}) will be dictated by the balance between several competing interactions. Firstly, the electrostatic contributions act in the sense of reducing the overall charge density inside the charged microgel, therefore increasing its size, and attempting to keep the counterions inside the microgel, in such a way as to neutralize its charge. The physical picture behind this is that similarly charged  monomers will try to be as far as possible from one another, leading to a stretching of the polymer chains. On the other hand, entropic effects make some counterions to leave the microgel, leaving space for the solvent molecules to come in. This contribution also tries to minimize the ionic density inside the microgel, and therefore leads to an increase of the particle size.  At the same time, solvent entropy tries to produce an uniform solvent distribution throughout the cell, leading to the penetration of solvent particles into the microgel. Again, this uptake of solvent molecules by the microgel produces the increase of the its volume. This effect is on the other hand counterbalanced by the repulsive solvent-polymer interactions, which will try to expel the solvent out from the microgel, decreasing its size. Finally, there is the elastic penalty for stretching the network, always trying to bring the microgel back to its unstressed state. Starting from the minimum volume (close-packing)  microgel configuration, the strong electrostatic forces between the mobile counterions will stretch the polymer network, while entropic effects will make solvent molecules to penetrate into the microgel. At some point, however, these effects are exactly counterbalanced by the elastic penalty for further increasing the particle size, together with the hydrophobic polymer-solvent repulsion effects. This is precisely when the Helmholtz free energy attain its minimum,  and therefore Eq. (\ref{equ}) is verified. 

Eq. (\ref{equ}) implicitly contains all the contributions to the microgel osmotic pressure. In particular the ionic contributions can, according to Eq. (\ref{Fion1}), be split into entropic and electrostatic contributions. As demonstrated by Barrat {\it et. al.} \cite{Barrat1992} in the context of the Donnan equilibrium theory, the entropic contributions strongly dominate over the electrostatic ones. The same  entropic dominance has been also observed in the similar case of star-shaped polyelectrolytes \cite{Jusufi_prl2002,Jusufi2002}. Due to the strong ionic condensation, a similar behavior is expected in the present situation, where the ionic contributions are obtained in the framework of the PB theory. It is important to note, however, that the mechanisms behind the ionic chemical equilibrium across the microgel interface will be slightly different. In the case of the ideal Donnan theory, chemical equilibrium is established between the condensed counterions and an infinite salt reservoir of concentration $c_{s}$. In the present case, the particle exchange across the microgel-solution interface must be such that the total number of particles inside the WS cell is conserved according to Eq. (\ref{RHOpm1}). At large microgel volume fractions, we expect this effect to play a non trivial role in the ionic chemical equilibrium.

\subsection{Numerical implementation}

Having established the theoretical basis of the model, we now turn to a short description of its numeral implementation. Due to the singular behavior of the PB equation, Eq.  (\ref{PB1}), close to the origin, a direct numerical integration of this equation is plagued by the numerical instabilities in this region -- particularly in regimes of highly charged microgels. The easiest way to avoid such instabilities is to rewrite Eq. (\ref{PB1}) as an integral equation for the electric field. 
Application of Gauss' Law, together with the spherical symmetry inside the cell allow us to transform the Eq. (\ref{PB1}) into
\begin{equation}
E(r)=\dfrac{\lambda_{B}Z}{r^2 a^3}(a^3-r^3)\Theta(a-r)-\dfrac{4\pi\lambda_{B}}{r^2}\int_{r}^{R}r'^{2}dr'\left(c_{+}e^{-\int_{r'}^{R}E(r'')dr''}+c_{-}e^{\int_{r'}^{R}E(r'')dr''}\right),
\label{PBE}
\end{equation}
where $E(r)\equiv \beta q \psi'(r)=\phi'(r)$ is the reduced electric field inside the cell, with the charge neutrality requiring that $E(R)=0$. The first term on the right-hand side of this expression represents the contribution to the electric field provided by the homogeneous monomer charge distribution, while the second term accounts for the inhomogeneous ionic distribution inside the cell. The coefficients $c_{\pm}$ are determined from the equilibrium distributions $\rho_{\pm}(r)=c_{\pm}e^{\mp\phi(r)}$, together with the requirement (\ref{RHOpm1}) of fixed number of ions inside the cell,
\begin{equation}
c_{\pm}=\dfrac{N_{\pm}}{4\pi\int_{0}^{R} {r^2 dr\exp\left(\mp \int_{r}^{R}E(r')dr'\right)}},
\label{cpm}
\end{equation}
and are themselves functionals of the electric field. It is easy to check that the solutions of Eqs. (\ref{PBE}) and (\ref{cpm}) automatically satisfy the desired boundary condition $E(0)=0$. 

The set of equations (\ref{PBE}) and (\ref{cpm}) have to be solved in a self-consistent fashion. This can be done by a direct Picard-like iteration procedure: starting from a guess field $E_{0}(r)$, the right-hand side of Eq. (\ref{PBE}) is numerically evaluated, allowing for the calculation of the output function $E(r)$. A new estimation for the electric field is then constructed by taking a proper combination of input and output fields, and the procedure is repeated until convergence is achieved. In most of the cases, however, this direct iteration procedure is unstable, resulting in non-convergent solutions. In order to stabilize the iteration procedure, a suitable combination of several previous input functions had to be taken. The coefficients for this combination are conveniently calculated according to the minimization criteria described by Ng \cite{Ng1974}, which strongly optimize numerical convergence. In the high charge regimes, up to $50$ coefficients had to be taken at each iteration step in order to achieve convergence. The numerical accuracy of the calculated ionic free energies was established by checking the equality of the osmotic pressure as calculated from the numerical derivative of Eq. (\ref{Fion1}) with respect to the cell volume, with the one resulting from the application of the contact value theorem at the cell edge \cite{Marcus1955}, to a reasonable degree of accuracy.

\section{Results}

Following Ref. [\onlinecite{Levin_Diehl2002}], we consider microgels comprising a total of $Nm=3\times10^7$ monomers of radius $r_{m}=3.2$\AA~ each, along with ions of radius $r_{i}=2$\AA. The Bjerrum length is set to be $\lambda_{B}=7.2$\AA, which is the typical value for an aqueous solution at room temperature. The WS cell radius is $R=2 \mu$m, corresponding to a concentrated microgel solution with overall density $\rho=0.03 \mu$m$^{-3}$. Two different situations are considered: $N=3\times10^5$ chains with monomer number $m=100$ and $N=6 \times 10^4$ polymer chains carrying $m=500$ monomers each. Since the product $mN$ is the same in both cases, the larger number of chains $N=3\times10^5$ corresponds, according to Eq. (\ref{Fel}), to a weaker deformability, whereas in the situation where the number of chains is smaller ($N=6 \times 10^4$), the microgels are more easily deformed.  We are now going to analyze separately three different aspects of this system, namely its swelling properties, the density profiles, and the effective charge of the microgel particles.

\subsection{Swelling}

\begin{figure}[h]
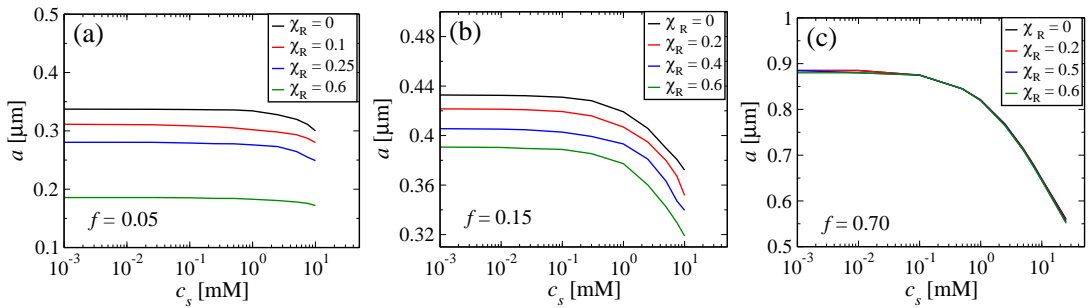

\centering
\includegraphics[scale=0.45]{fig2a.eps}
\includegraphics[scale=0.45]{fig2b.eps}
\includegraphics[scale=0.45]{fig2c.eps}
\caption{Microgel radius $a$ as a function of the salt concentration $c_{s}$ for different Flory parameters $\chi_{R}=\chi(r_{i}/r_{m})^3$, and fractions of dissociation $f=0.05$ (a), $f=0.15$ (b) and $f=0.7$ (c). The total number of polymer chains is $N=3\times10^5$, each one carrying an average of $m=100$ monomers.}
\label{fig1} 
\end{figure}

Fig. \ref{fig1} shows the effects of salt on the swelling process, in the case of microgels with $N=3\times 10^5$ chains, for different renormalized Flory parameters $\chi_{R}\equiv \chi(r_{i}/r_{m})^3$ and different dissociation fractions $f$. Clearly, the increase in salt concentration beyond certain amount leads to a considerable reduction in the particle size in the regime of sufficiently high microgel charges. This is a consequence of the strong ionic imbalance across the microgel-solution interface, which results in an increase of the pressure exerted by the external ions on the microgel surface (Donnan effect). For weakly charged microgels ($f=0.05$), addition of salt has a minor effect on the particle size. In this regime of weak electrostatic coupling, the swelling is strongly dominated by the solvent contributions. As the Flory parameter $\chi_{R}$ is increased, solvent molecules are expelled out from the microgel, resulting in a reduction of the particle size. In the case of moderate microgel charge ($f=0.15$), the swelling is influenced by both solvent and electrostatic contributions, and the particle size is considerably reduced when the salt concentration increases beyond a certain value. In the opposite limit of strongly charged microgels, electrostatic effects start to dominate over the solvent interactions, and the particle size becomes very weakly dependent on $\chi_{R}$, as shown in Fig. \ref{fig1}c for the case $f=0.7$. The particle shrinking with the addition of salt is however more pronounced at larger dissociation fractions $f$. In the case $f=0.7$, the microgel volume becomes approximately $4$ times smaller as the salt concentration is increased from $c_{s}=10^{-3}$ mM to $c_{s}=10^{-1}$ mM.  Again, this effect can be easily understood in terms of the corresponding stronger ionic discontinuity across the microgel-solution interface for larger fractions $f$.

\begin{figure}[h]
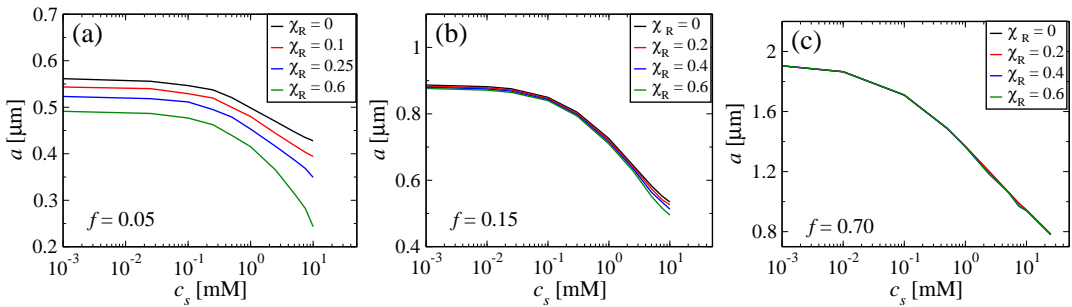

\centering
\includegraphics[scale=0.45]{fig3a.eps}
\includegraphics[scale=0.45]{fig3b.eps}
\includegraphics[scale=0.45]{fig3c.eps}
\caption{Same as in Fig. \ref{fig1}, but with $N=6\times10^4$ polymer chains and $m=500$ monomers per chain.}
\label{fig2} 
\end{figure}

A similar scenario is observed in the case of microgels with a smaller number of chains ($N=6\times 10^4$), as is shown in Fig. \ref{fig2}. The effect of salinity, however, is enhanced in this case. Since the elastic penalty is reduced (see Eq. \ref{Fel}), the microgels are easily deformed by the external pressure, so that addition of salt has a stronger effect. Furthermore, the electrostatic effects become dominant even in the case of moderate charged microgels ($f=0.15$), where the microgel size becomes very weakly dependent on $\chi_{R}$. Since the particle is more easily expanded is this case, there is an entropic gain from the incoming solvent molecules, which overcomes the  solvent-monomer repulsion.  Even in the case of weakly charged microgels ($f=0.05$), the ionic contributions play an important role, and the particle size is considerably reduced when the salt concentration increases. In the case of strongly charged microgels $f=0.7$, the microgel volume becomes now about $8$ times smaller when the salt concentration is increased from $c_{s}=10^{-3}$ mM to $c_{s}=10^{-1}$ mM.

In all the situations, the microgel radius changes very slowly at small salt concentrations. However, as the amount of added salt grows beyond some value ($c_{s}\approx 0.01$ mM for $N=6\times 10^4$ and $c_{s}\approx 0.1$ mM for  $N=3\times 10^5$), an abrupt decay of the microgel size is observed. This dramatic reduction in the particle size with the increase of salt concentration after a certain limit is also predicted by the traditional Donnan and Debye-H{\"u}ckel theories \cite{Quesada-Perez2011,Rydzewski1990,Victorov2006,Yigit2012}, and has been strongly supported by experimental measurements \cite{Katchalsky1955,Fernandez-Nieves2001,Capriles-Gonzalvez2008,Nisato1998,Hooper1990,Lopez-Leon2006,English1996}.  In Ref. [\onlinecite{Lopez-Leon2007}] it was experimentally shown that a similar behaviour also holds  for the case of addition of salt in neutral microgels. Due to the absence of Donnan effect \cite{Donnan1924} in the case of neutral microgels, the region of de-swelling in that case is shifted to higher salt concentrations ($c_{s} \approx 100$ mM), where then the entropic contributions from the ions become overwhelmingly dominant \cite{Lopez-Leon2007}.

\begin{figure}[h!]
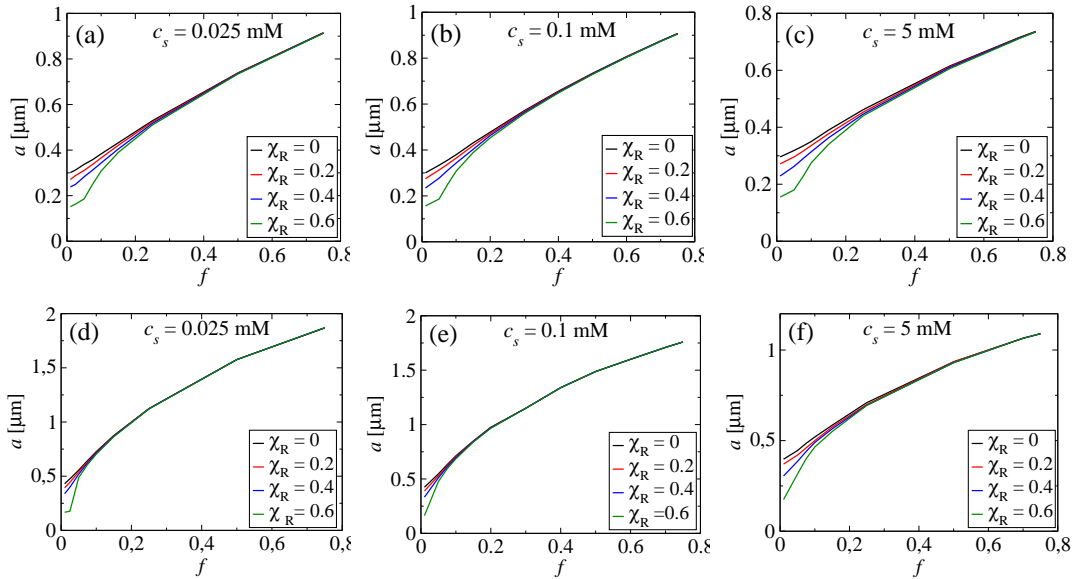

\centering
\includegraphics[scale=0.425]{fig4a.eps}
\includegraphics[scale=0.425]{fig4b.eps}
\includegraphics[scale=0.425]{fig4c.eps}\vspace{0.2cm}\\
\includegraphics[scale=0.41]{fig4d.eps}
\includegraphics[scale=0.41]{fig4e.eps}
\includegraphics[scale=0.41]{fig4f.eps}
\caption{Microgel radius $a$ as a function of the fraction of dissociated monomers $f$ for different Flory parameters $\chi_{R}=\chi(r_{i}/r_{m})^3$ and salt concentration $c_{s}=0.025$ mM (a and d), $c_{s}=0.1$ mM (b and e) and $c_{s}=5$ mM (c and f). The curves (a), (b) and (c) represent the case ($N=3\times 10^5$), while curves (d), (e) and (f) corresponds to microgels with the lower number of chains ($N=3\times 10^4$).}
\label{fig3} 
\end{figure}

For the two cases of $N$ considered, the microgel swelling appears to be strongly influenced by the dissociation fraction $f$. For highly charged microgels, the swelling is mostly dictated by the ionic contributions, whereas for smaller fractions $f$ the solvent interactions start to play an important role, and the particle size becomes highly $\chi$-dependent. This trend is verified in Fig. \ref{fig3}, where the microgel size as a function of $f$ for different salt concentration and Flory parameters is shown. In all the cases, the particle radius $a$ increases significantly as the microgel charge grows larger. By increasing the degree of ionic dissociation, the electrostatic repulsion between the charged backbones becomes stronger, resulting in an expansion of the polymer chains. Again, these electrostatic effects become stronger in the case where the the polymer network is more flexible ($N=6\times 10^4$). 

\begin{figure}[h]
\centering
\includegraphics[width=7.5cm]{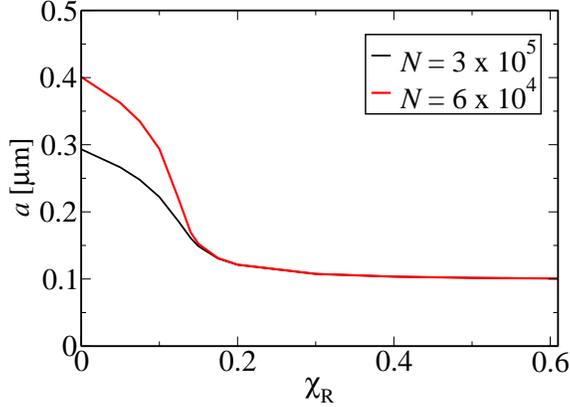}
\caption{Microgel radius $a$ in the limit of an uncharged microgel as a function of the renormalized Flory parameter $\chi_{R}$ corresponding to microgels with  $N=3\times10^5$ polymer chains (black curves) and $N=6\times10^4$. The salt concentration if fixed in $c_{s}=0.01$ mM. The particle size is obtained by numerically solving  Eq. (\ref{nocharge}). }
\label{fig8}
\end{figure}

As Fig. \ref{fig3} suggests, the effects from the ionic Donnan equilibrium become negligible in the case of small fraction of dissociated ions. Moreover, the particle size is not influenced by the increase in salt concentration is this limit, as can be seen in Fig. \ref{fig1}a. The swelling behavior will be therefore mostly dictated by the polymer-solvent interactions when $f$ is small enough. In fact, when $f\ll 1$ the ionic concentrations become approximately constant throughout the cell, and the equilibrium condition Eq. (\ref{equ}) reduces to
\begin{eqnarray}
\left(1-\dfrac{8\pi c_s r_{i}^3 }{3} \right)\ln\left[1-\dfrac{3Nmr_{m}^3}{a^3(3-8\pi c_s r_{i}^3)}\right]+Nm\left(\dfrac{r_m}{r_i}\right)^3\left(1+\dfrac{\chi_{R} mNr_{m}^3}{a^3}\right) \nonumber \\
+\dfrac{N}{2}\left(\dfrac{2 a^2}{N^{2/3}m^{2/3}r_{m}^{2}}-1\right)=0.
\label{nocharge}
\end{eqnarray}
Since $c_{s}r_{i}^3\ll 1$ for all the experimentally relevant salt concentrations, it results from this relation that the microgel size depends very little on the amount of added salt in the limit of neutral polymer networks, in accordance with Fig. \ref{fig1}a. The particle radius resulting from this equation are shown in Fig. (\ref{fig8}) for a salt concentration $c_{s}=0.025$ mM, and two different number of chains. For $\chi_{R}> 0.2$, the radius $a$ becomes independent of both Flory parameter and the number of chains in the polymer network.

\subsection{Ionic Profiles}

\begin{figure}[h]
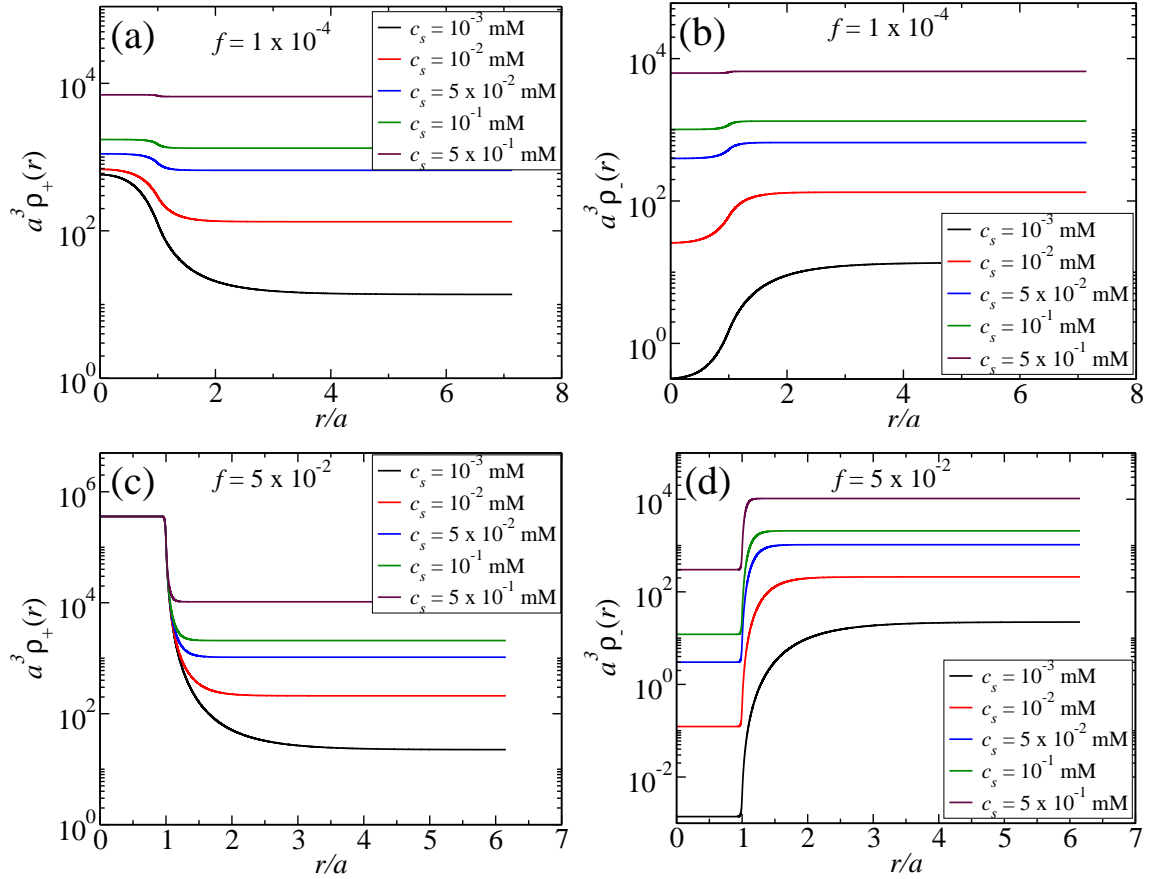

\centering
\includegraphics[width=7.5cm]{fig6a.eps}
\includegraphics[width=7.5cm]{fig6b.eps}\vspace{0.2cm}\\
\includegraphics[width=7.5cm]{fig6c.eps}
\includegraphics[width=7.5cm]{fig6d.eps}
\caption{Ionic density profiles obtained at different salt concentrations, corresponding $\chi_{R}=0.1$ and a number of $N=3\times 10^5$ chains. The microgels are negatively charged, and the fractions of dissociated monomers are $f=10^{-4}$ (a and b) and $f=0.05$ (c and d).  For visualization proposes, the $y$ axis is displayed on the logarithmic scale.}
\label{fig4}
\end{figure}

For a given set of system parameters, the equilibrium ionic density profiles are the ones that satisfy both the PB equation, Eq. (\ref{PB1}), and the equilibrium condition, Eq. (\ref{equ}), simultaneously. The resulting density profiles for the case $\chi_{R}=0.1$ and $N=3\times 10^5$ are shown in Fig. \ref{fig4}, for several different salt concentrations and fractions of dissociation $f=10^{-4}$ (Figs. \ref{fig4}a and \ref{fig4}b) and $f=0.05$ (Figs. \ref{fig4}c and \ref{fig4}d). Due to strong electrostatic interactions, the density distributions are highly inhomogeneous across the microgel-solution interface. While the counterions are accumulated inside the microgel, the coions are expelled out of this region. This abrupt change of ionic concentrations resulting from the charge balance across an interface is followed by a strong electric field gradient, and is known in the chemical-physics literature as the Donnan effect \cite{Tamashiro1998,Donnan1924}. This effect is more pronounced at small salt concentrations, where electrostatic effects clearly dominate \cite{Chepelianskii2009,Bauli2012}. As the salt concentration increases, the entropic contributions start to rival the electrostatic ones, resulting in more homogeneous ionic distributions \cite{Bauli2012}.  For the same reason, it also becomes favorable for the coions to penetrate the microgel, as can be clearly seen in Fig. \ref{fig4}b. While the ionic inhomogeneities at the microgel surface are smooth for weakly charged microgels ($f=10^{-4}$), the profiles become very sharp already at moderate charged macroions ($f=0.05$). In this case, the electrostatic coupling is so strong that the counterion penetration is approximately the same for all the salinities considered. Inside the microgel, a \textit{local} charge neutrality is achieved (zero electric field), resulting in almost uniform distribution functions (Figs. \ref{fig4}c and \ref{fig4}d). This is followed by a strong electric field difference across the interface, responsible for the charge gradient in this region.  Clearly, the uniform pattern observed in the microgel interior is a consequence of the homogeneous charge distribution assigned to the microgel charge, Eq. (\ref{rhom}). The local charge neutrality resulting from the PB equation confirms the charge-neutral picture that has been assumed in experiments as well as in many theoretical models for the ionic contributions \cite{Quesada-Perez2011}.

\subsection{Effective and renormalized charges}

\begin{figure}[h]
\centering
\includegraphics[width=8cm]{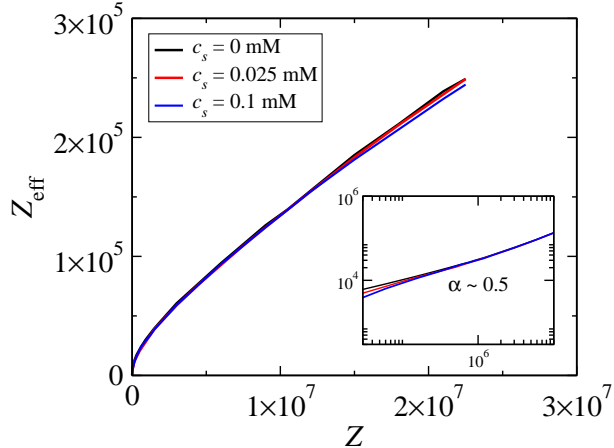}
\caption{Effective microgel charge as a function of the polymer charge, for different salt concentrations. The renormalized Flory parameter is $\chi_{R}=0.1$, while the number of polymer chain is in this case $N=3\times 10^5$. The inset shows the same curves in a double-logarithmic scale at large values of $Z$.}
\label{fig5}
\end{figure}

Due to the strong electrostatic interaction, most of the dissociated counterions  remain trapped inside the polymer backbones, as can be clearly seen in Figs. \ref{fig4}a and \ref{fig4}c. As a consequence, the microgels have an effective net charge $Z_{\mathrm{eff}}$, whose magnitude is much smaller than the initial polymer charge, $Z_{\mathrm{eff}}\ll Z$. This ionic penetration effect has been experimentally verified through electrophoretic mobility measurements \cite{Finessi2011,Lopez-Leon2006,Andersson2006}. Within the present model, the effective charge can be easily obtained as a functional of the calculated density profiles
\begin{equation}
Z_{\mathrm{eff}}=Z-4\pi\int_{0}^{a}r^2 \left[\rho_{+}(r)-\rho_{-}(r)\right]dr=\dfrac{a^2 E(a)}{\lambda_{B}},
\label{Zeff}
\end{equation}  
where $E(a)=\phi'(a)$ is the reduced electric field at the microgel surface. The typical behavior of this quantity as a function of the bare polymer charge for three different salt concentrations is displayed in Fig. \ref{fig5}. For a given set of parameters, the effective charge corresponding to high polymer charges $Z$ shows a weak dependence on the amount of added salt. This result is consistent with the calculated density profiles (see Fig. \ref{fig4}a), where the counterion condensation is practically the same for a wide range of salt concentrations.  At large values of $Z$ the effective charge increases monotonically, showing a perfect power-low dependency $Z_{\mathrm{eff}}\sim Z^{1/2}$  (inset of Fig. \ref{fig5}). This behavior is quite general and holds in fact for arbitrary particle sizes, as has been analytically demonstrated by Chepelianskii {\it et al} in the context of the PB equation for penetrable macroions in salt-free solutions \cite{Chepelianskii2009,Chepelianskii2011}, and further extended by Bauli {\it et al} for the case of added salt \cite{Bauli2012}.  The same scaling law for the effective charge has been also observed experimentally \cite{Levin_Diehl2002}.

In the case of microgel systems, a clear distinction must be made between the aforementioned effective charge and the concept of \textit{charge renormalization}, usually employed in the description of hard colloids \cite{Levin2002}. While the effective charge represents the net microgel charge -- which accounts for the ionic penetration -- the renormalized charge is an effective parameter designed to incorporate  the strong non-linear effects resulting from the large charge asymmetry between macroions and the small ions \cite{Chepelianskii2009,Bauli2012}.  When dealing with linear theories for describing highly charged systems (e.g. Yukawa-like models), it is the renormalized charge which should be used as input to implicitly account for non-linear effects. 

In the framework of the mean-field cell model, the renormalized charge can be easily obtained through the so-called \textit{Alexander prescription}, which has been successfully employed in the case of hard colloidal systems \cite{Alexander1984,Trizac2003}, and recently extended to account for counterion penetration \cite{Chepelianskii2009,Bauli2012}. The basic idea is to linearize the PB equation, Eq. ({\ref{PB1}}), around the potential at edge of the WS cell. The resulting  potential $\phi_{\mathrm{lin}}(r)$ satisfies the following liner equation, 
\begin{equation}
\nabla^2 \phi_{\mathrm{lin}}(r)=\kappa^2 (\phi_{\mathrm{lin}}(r)-\phi_{R})-4\pi\lambda_{B}\left(\tilde{\rho}_{+}-\tilde{\rho}_{-}-\dfrac{3Z_{\mathrm{ren}}}{4\pi a^3}\Theta(a-r)\right),
\label{PBL}
\end{equation}
where $\phi_{R}\equiv\phi(R)$ is the potential at the cell boundary, $\tilde{\rho}_{\pm}=c_{\pm}e^{\mp \phi_{R}}$ are the corresponding ionic densities at $r=R$, and $\kappa \equiv \sqrt{4\pi\lambda_{B}(\tilde{\rho}_{+}+\tilde{\rho}_{-})}$ is the inverse of the (renormalized) screening length. Note that we have replaced
$Z$ by $Z_{\mathrm{ren}}$.  
By extending the linear solution $\phi_{\mathrm{lin}}(r)$ throughout the cell volume, the macroion charge must assume a different value  from the bare charge $Z_{\mathrm{ren}}\ll Z$, in order
to produce the asymptotically correct potential and electric field at the cell boundary \cite{Trizac2003}. The calculation is done as follows. For a given set of parameters, Eq. (\ref{PB1}) is solved numerically, and the electric potential at the cell border $\phi_R$ is calculated. This potential is then used as an input in Eq. (\ref{PBL}), which is solved under the boundary conditions $\phi_{\mathrm{lin}}(R)=\phi_R$ and $\phi'_{\mathrm{lin}}(R)=0$. These conditions guarantee that both linear and non-linear solutions provide the same electrostatic potential and electric field at the cell boundary.  Eq. (\ref{PBL}) can be solved analytically, resulting in the linear potential
\begin{eqnarray}
\phi_{\mathrm{lin}}(r)&=-\gamma\left[\dfrac{(\kappa^2 Ra-1)\sinh(\kappa(a-R))+\kappa(a-R)\cosh(\kappa(a-R))}{\kappa a \cosh(\kappa a )-\sinh(\kappa a)}
\dfrac{\sinh(\kappa r)}{\kappa r}-1\right]\nonumber \\\nonumber\\
&-\dfrac{3Z_{\mathrm{ren}}\lambda_{B}}{\kappa^2 a^3}+\phi_{R},
\label{phi_lin1}
\end{eqnarray}
for $r\le a$,  and 
\begin{equation}
\phi_\mathrm{{lin}}(r)=-\dfrac{\gamma}{\kappa r}\left[\kappa R\cosh(\kappa(r-R))+\sinh(\kappa(r-R))\right]+\gamma+\phi_{R},
\label{phi_lin2}
\end{equation}
for $a< r\le R$, where $\gamma \equiv 4\pi\lambda_{B}(\tilde{\rho}_{+}-\tilde{\rho}_{-})/\kappa^2$. The renormalized charge in Eq. (\ref{phi_lin1}) follows from the requirement that Eqs. (\ref{phi_lin1}) and (\ref{phi_lin2}) must be equal at $r=a$. Applying this condition, the renormalized charge can then be written as
\begin{equation}
Z_{\mathrm{ren}}=\dfrac{\gamma\kappa^2 a ^3}{3\lambda_{B}}\left[\dfrac{\left(\kappa R-\tanh(\kappa a)\right)\cosh(\kappa(a-R))+\left(1-\kappa R\tanh(\kappa a)\right)\sinh(\kappa(a-R))}{\kappa a -\tanh(\kappa a)}\right].
\label{Zren}
\end{equation}
For a given microgel radius $a$ and salt concentration $c_{s}$, the only input necessary for the calculation of $Z_{\mathrm{ren}}$ are the potential $\phi(R)$ and concentrations $\tilde{\rho}_{\pm}$, which follow directly from the nonlinear solution of the PB equation. 

\begin{figure}[h]
\centering
\includegraphics[width=8.5cm]{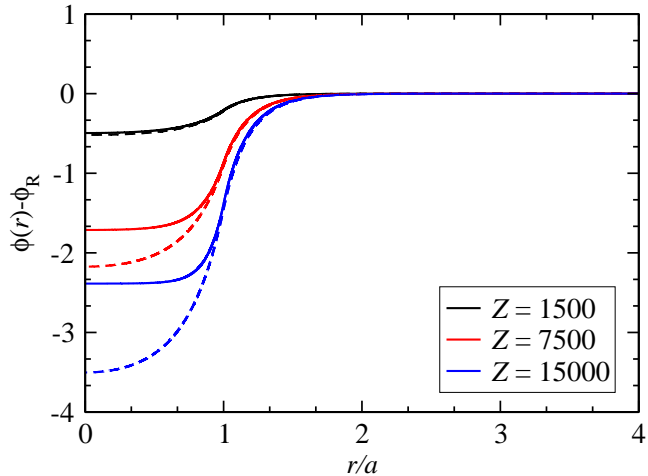}
\caption{Comparison between linear (dashed lines) and non-linear (solid curves) solutions of the PB equation, Eqs. (\ref{PB1}) and (\ref{PBL}) respectively, for different microgel charges. In the linear solution, the bare charge $Z$ has been replaced by the renormalized one, $Z_{\mathrm{ren}}$. The salt concentration is $c_{s}=0.025$ mM, the Flory parameter was set at  $\chi_{R}=0.1$, and the number of polymer chains is $N=3\times 10^5$.  }
\label{fig6}
\end{figure}

In Fig. \ref{fig6}, the linear potential from Eqs. (\ref{phi_lin1}) and (\ref{phi_lin2}) is compared with the numerical solutions of the PB equation, Eq. (\ref{PB1}). In the linear regime of moderate microgel charges ($Z\approx 1500$), the linear solution reproduces quite well the non-linear potential, and the renormalized charge is approximately equal to the bare polymer charge, $Z_{\mathrm{ren}}\approx Z$. As the microgel charge increases, the non-linear effects become progressively more relevant and strong deviations between the linear and the non-linear solutions are observed at small distances from the microgel center. The renormalized charge, however, ensures that the linear solution is able to correctly account for the large-distance behavior of the non-linear potential.

\begin{figure}[h]
\centering
\includegraphics[width=8.5cm]{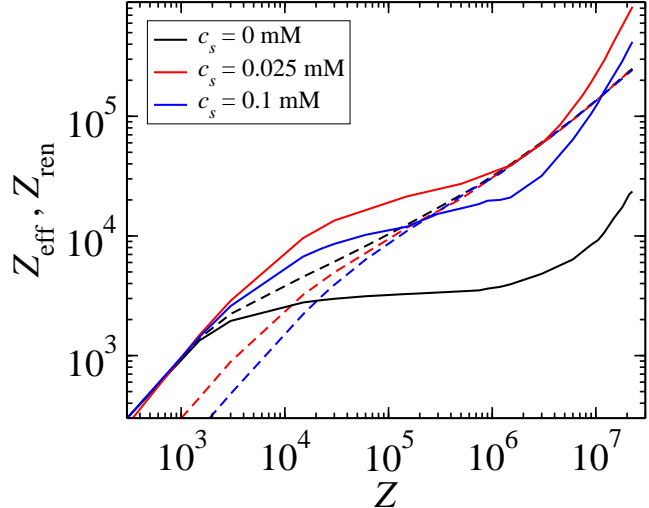}
\caption{Comparison between the effective (dashed lines) and the renormalized charges (continuous lines), for a microgel solution with $\chi_{R}=0.1$. The curves are shown in a double-logarithmic scale. Black curves represent the absence of salt ($c_{s}=0$), red lines stand for the case with $c_{s}=0.025$ mM, while the blue ones correspond to $c_{s}=0.1$ mM.}
\label{fig7}
\end{figure}

In Fig. \ref{fig7} the renormalized charge resulting from Eq. (\ref{Zren}) (solid curves) is compared with the effective one (dotted curves), Eq. (\ref{Zeff}), as a function of the bare microgel charge, for different salt concentrations, 
Flory parameter $\chi_{R}=0.1$ and  $N=3\times 10^5$. Clearly, the qualitative behavior of these quantities as functions of the bare polymer charge is completely different. Upon addition of salt, the ionic condensation is already present even in the case of small microgel charges, resulting in an effective charge much smaller than the microgel charge. The renormalized charge, in contrast, coincides with the bare polymer charge in this linear limit, $Z_{\mathrm{ren}}\approx Z$, as can be clearly identified by the linear curves with slope $1$ close to the origin. A similar linear relation is also observed for the effective charges in the absence of salt (dotted black curve). Clearly, the additional of salt makes it favorable for the counterions to penetrate the microgel, leading the strong counterion condensation even at small microgel bare charges. As the bare microgel charge increases beyond this linear regime, quite different functional behaviors for the effective and renormalized charges are observed: while the former grows as a power-law for large values of $Z$ (see Fig. \ref{fig5}), the renormalized charge increases much faster at larger bare charges $Z$, in a way that clearly deviates from the simple power-low trend. This qualitative behavior is quite distinct from the classical picture observed in the case of hard colloids in the presence of monovalent ions, where the renormalized charge rapidly achieves a saturation value beyond the linear regime \cite{Trizac2003}. In the case of charged microgels, the particle swelling drived by the increase of the microgel bare charge (see Fig. \ref{fig3}) prevents this saturation regime to be reached. Instead, the renormalized charge grows monotonically as the microgel charge (and therefore the particle size) grows further. 

Due to the large difference observed between the effective and renormalized charges, it is extremely important to rely on the renormalized charge (instead of $Z_{\mathrm{eff}}$) as the relevant input parameter in order to properly account for strongly non-linear effects, while describing thermodynamic and structural properties of highly charged microgels through the traditional, Yukawa-like theories. In the limit of relatively small polymer charges, the linear theory is quite accurate, and the bare polymer charge is sufficient to correctly account for the system properties. Analogous conclusions have been recently reported for the similar case of hydrophobic polyelectrolytes \cite{Chepelianskii2009}, as well as for core-shell like charged polymers in the presence of monovalent salt \cite{Bauli2012}.

\section{Conclusions}
  
A simple model has been put forward to calculate the equilibrium properties of charged microgels in the framework of the traditional PB and Flory theories. Particular emphasis was given to the role of salt. While the effective charges are weakly influenced by the addition of salt, a strong salt dependence was found for the swelling behavior, the renormalized charges, as well as for the ionic density distributions. The Alexander prescription for charge renormalization \cite{Alexander1984} was extended to the situation of penetrable macroions with varying particle size. It was shown that the effective and the renormalized charges behave dramatically differently in the regime of high microgel charges.

For highly charged microgels, the calculated ionic profiles show a very simple functional behavior, in which both coion and counterion distributions are approximately uniform inside the microgel, resulting in a local charge neutrality. This behavior is clearly a consequence of the uniform charge distribution assigned to the charged backbones \cite{Moncho_Anta2013,Chepelianskii2009,Bauli2012,Barrat1992}. It is known, however, that the highly inhomogeneous counterion distribution of the trapped counterions might have a strong influence on the resulting swelling behavior \cite{Jusufi2002,Moncho_Anta2013,Moncho2013}. A possible improvement of the theory will be to consider non-uniform monomer distributions inside the microgel. Another limitation of the present model is the absence of the counterion-polymer steric interaction: when calculating the ionic free energy in Eq. (\ref{Fion1}), it is assumed that the ions are free to move throughout the WS cell. It is well known, however, that the mobility of the counterions is dramatically reduced by their strong electrostatic interaction with the microgel backbone. A proper way to account for this entropic limitation is to explicitly consider the exclusion volume polymer-ion interaction. Once a distribution is assigned to the polymer chains, these steric effects can be incorporated with the formulation of a weight-density functional theory, in the framework of the Rosenfeld fundamental measure theory \cite{Rosenfeld1996,Frydel2012}.  Work along these lines is currently in progress.            

\section{Acknowledgments}
This work was partially supported by the CNPq, INCT-FCx, and by the US-AFOSR under the grant 
FA9550-12-1-0438.


%

\end{document}